	\documentclass{article}[11pt]

	\usepackage{microtype} 
	\usepackage{enumitem} 
	\setlist{nolistsep,leftmargin=*}
	
	\usepackage{xspace} 

	\usepackage{booktabs}

	\usepackage{soul} 
	\sodef\myspace{}{.09em}{.6em plus.1em minus.1em}{.6em plus.1em minus.1em}
	\sodef\mysmallspace{}{.07em}{.6em plus.1em minus.1em}{.6em plus.1em minus.1em}

	\usepackage{cinzel} 
	\usepackage{pacioli} 
	\usepackage{tgpagella}
	\usepackage{newpxmath} 
	\usepackage{newpxtext} 
	\usepackage{hologo} 
	\usepackage{amssymb}

	\usepackage{amsmath}  

	\usepackage{units} 
	\usepackage{siunitx} 
	\usepackage{miller}

	\def\innervector(#1,#2, #3, #4){\ensuremath{\left( #1,#2, #3, #4 \right)}}

	\usepackage{xcolor} 
	\definecolor{Maroon}{HTML}{800000} 
	\usepackage{tcolorbox} 

	\usepackage{graphicx} 
	\setkeys{Gin}{width=\linewidth,totalheight=\textheight,keepaspectratio} 
	\usepackage{float}

	\usepackage{geometry}
	\usepackage{multicol} 
	\usepackage{lettrine} 

	\usepackage{titlesec} 
	\titleformat{\section}
	{\bfseries\Large\color{Maroon}}
	{\thesection}{0.5em}{\MakeUppercase}
	
	\titleformat{\subsection}
	{\bfseries\large\color{black}}
	{\thesubsection}{0.5em}{}

	\usepackage{cite}
		
	\let\OLDthebibliography\thebibliography
	\renewcommand\thebibliography[1]{
	  \OLDthebibliography{#1}
	  \setlength{\parskip}{0pt}
	  \setlength{\itemsep}{0pt plus 0.3ex}
	}

\renewenvironment{abstract}{
\begin{center}
\bfseries
\end{center}}

\usepackage[affil-it]{authblk}

\usepackage{fancyhdr}
\begin{document}


\title{\bfseries{\huge Version 0: An Educational Package for Helium Atom Scattering Studies  }}
\author{\bfseries\large 
E.L. Arnold, M-S. Liu, R. Prabhu,  C.S. Richards, D. Ward, N. Avidor
\footnote{Correspondance to all authors to atomscattering@phy.cam.ac.uk}
 }
\affil{\bfseries\large\textrm{
Cavendish Laboratory, University of Cambridge
}}
\date{\bfseries\large 
24th September 2020
}

\maketitle


\renewcommand{\thefootnote}{\roman{footnote}}

\begin{abstract}
\vspace{-30pt}
\begin{tcolorbox}[colback=red!5!white,colframe=Maroon, arc = 0pt, boxrule = 1pt, title = {\center \large \textbf{Abstract}}]

\noindent
Helium atom scattering studies have the potential for making numerous breakthroughs in the study of processes on surfaces. As this field remains active, there will frequently be new young researchers entering the field. The transition from student to researcher is often met with difficulty, consequently wasting limited time available for a PhD or master's level research. Addressing this issue, we present an educational package for emerging research students in the field of helium atom scattering. We hope that this package serves as sufficient material to significantly accelerate the progress made by new postgraduate students.

\end{tcolorbox}
\end{abstract}

\begin{multicols}{2} 

\section{Introduction}
Helium atom scattering has remained a fruitful field throughout the time from its inception in the 1980s \cite{PhysRevLett.44.1417, PhysRevLett.46.437, PhysRevB.27.3662} to the present day. Consequently, there has been significant interest in the improvement of existing techniques during this time period, and the development of novel machines; this promotes the encouragement of increasingly more sensitive measurements to be made of surface processes on solids. One example of this is the development of a sophisticated helium-3 spin-echo ($^3$HeSE) spectrometer \cite{JARDINE2009323} by the Cambridge group, building on a prototype from \cite{PhysRevLett.75.1919} in the 1990s. This machine has been one of the main research focusses for the Cambridge group for over a decade; thus it is highly important new research students have access to high quality resources to introduce its use. In the light of the importance of spin-echo to the group, a significant proportion of the resources presented in this article are oriented towards building an understanding of the spin echo machine.

In this document, we discuss a package created for a Summer internship programme at the Surfaces group in Cambridge. In section 2, we discuss the Educational Package itself (which constitutes a significant bulk of the programme); section 3 focusses on the additional activities undertook by the students; section 4 discusses the future ambitions in expanding the Educational Package to serve an even greater selection of topics, and section 5 briefly concludes the content of the current article.

\section{Educational Package}
The educational package \cite{Edu.Package} presented in this article contains two main features. The focus of this package is 15 so-called "assignments", which are computational investigations into a particular area of relevance in surface physics. These are designed in such a way to partially bridge the gap from typical British undergraduate teaching to research. A list of these is given in the appendices. These "assignments" are accompanied by a handbook of the theory needed to complete them. We believe this handbook is written in such a way to promote the development of intuition early in the study of the relevant physics, especially the less rigorous intuition that is difficult to promote in ordinary research papers. It is styled to mimic typical undergraduate texts, while still being oriented towards research students. The package further mimics the convenient style of reading lecture notes followed by attempting a problem, which should be particularly familiar to most undergraduates.

\section{Summer Projects}
In the Summer of 2020, the Surfaces, Microstructure and Fracture group at Cambridge organised a (remote) set of computational research projects for undergraduates at the University of Cambridge. These were an ambitious set of projects related to the research output of the group and collaborators from Wales, Austria, Australia and the Open University; most of the projects expand on published work from both students and faculty.

At the time of writing, only the pilot year of the projects had been run, under the name of the Undergraduate Summer Research Program in Surface Nanophysics and Atom-Surface Scattering.  What did the projects in the pilot year consist of? The students completed research-style tasks from the prototype assignments, without abandoning the teaching style familiar to them. During the same time frame, regular seminars were held, allowing the relevant theory to be introduced in a more passive format. Once these had been completed, the students regularly met with an assigned project supervisor to plan and solve a research activity. Overall, the pilot year programme took most students 10 weeks to complete, of which two weeks were assigned to solving the prototype assignments. 

We hope that future years of the projects can rely heavily on the Educational Package presented as the main product of this article, which should greatly ease the first month's work in the projects. This package is designed to be flexible; thus we hope that it can be easily extended to support the progress of postgraduate students;  we further anticipate that it can support those outside of the surfaces group at Cambridge. The package does not require the support of the authors, or the faculty members running the programme, to be effective.

\section{Future Work}
The package presented in this document has been labelled as "version 0" rather than "version 1". This is because the process for internally checking and updating the package is anticipated to take a significant length of time (as it is approximately 300 pages in length!), so the package we present to date cannot be assured to be free of errors. We hope that the package will be rereleased in the future as a higher quality set of documents. The authors aim to submit an updated publication during late 2021.

Furthermore, we hope that documents pertaining to helium atom microscopy and time-of-flight methods can be added to the package, along with documents aimed to serve more general surface science.

\section{Conclusions}
The package we have written is aimed to ease the transition from lectured courses in physics to research-oriented studies in helium scattering. Elements of this package may find use in other fields, especially those related to atom scattering or surface diffusion. We hope that, in the future, this package may be expanded by other Summer interns in the surfaces group at the Cavendish Laboratory.

\section*{Acknowledgements}
The authors declare they have no financial interests related to the resources discussed in this article. 

E.L.A., M.-S.L., R.P., and C.S.R. issue special thanks to N.A. for organising the summer programme.

{\scriptsize
\bibliography{main} 
\bibliographystyle{naturemag}}

\section*{Appendices}

\subsection*{Appendix A: Location of Resources}
The resources discussed in this document are freely available at \cite{Edu.Package}. 

\subsection*{Appendix B: List of Resources (Educational Package)}
There are four categories of documents in the educational package. Summarily, these are the 15 assignments, their accompanying handbook, a readme file (\textit{i.\hspace{1pt}e.}\hspace{2pt}a brief manual for the package), and this paper. The titles of the assignments are:
\begin{enumerate}
\item Graph Plotting
\item Fourier Transforms
\item Hard Wall Potentials
\item The Eikonal Approximation
\item The Chudley-Elliot Model
\item Monte Carlo Simulations
\item Molecular Dynamics
\item The Intermediate Scattering Function
\item Adsorbates
\item The Wavelength Intensity Matrix
\item Bayesian Fitting
\item Spin Precession
\item Potential Energy Surfaces
\item Rotating Molecules
\item Radial Distribution Function
\end{enumerate}
In addition to the assignments, the chapters in the associated "Theory Handbook" are:
\begin{enumerate}
\item Fourier Transforms
\item Convolution
\item Crystal Structures
\item Verlet Integration
\item Langevins
\item Bulk Scattering
\item Surface Scattering
\item Adsorbates
\item Scattering Functions
\item Spin Echo
\item The Radial Distribution Function
\item Conclusions
\end{enumerate}


\end{multicols}

\end{document}